# WOMEN IN PHYSICS: SCISSORS EFFECT
## from the Brazilian Olympiad of Physics to Professional Life

Débora P. Menezes[1*], Carolina Brito[2*], Celia Anteneodo[3*]

(1) Departamento de Física - CFM – UFSC
(2) Instituto de Física, UFRGS
(3) Departamento de Física – PUC-Rio
(*) Members of Grupo de Trabalho sobre Questões de Gênero da SBF (GTG-SBF)

**Abstract**

Despite the legislation in force in most countries and the discourse on gender equality, the observed reality, still today, shows a clear picture of inequality and restrictions for women, particularly when the issue is the choice of a career and the subsequent promotions. In many professions, although women represent the majority in the entrance to undergraduate courses, their participation is dramatically reduced as the career progresses to the highest levels. While the participation of men increases, that of women decreases as the academic levels move towards the top ranks. This kind of evolution of two groups in opposed directions is usually referred as the "scissors effect". This effect with respect to sexes also occurs in the area of physics, both in Brazil and in the rest of the world. But at what age does this effect begin? What is its intensity in the different stages of women's lives? In this article we aim to shed some light on these questions by analyzing data of the Brazilian Olympiad of Physics and of fellowships for physicists in Brazil. We observe the systematic occurrence of scissors effect throughout the full studied data, ranging from the first years of participation in the Olympiads (13 years old) to the highest stages of the scientific career.

**Efeito tesoura**

Apesar das leis vigentes na maioria dos países e do discurso de igualdade entre gêneros, a realidade observada, ainda nos dias de hoje, mostra um quadro desigual e limitador para as mulheres, especialmente quando o assunto é escolha da carreira e posterior ascensão.

Em diversas profissões, a participação das mulheres, mesmo sendo em alguns casos majoritária no ingresso aos cursos de graduação, vai se reduzindo notadamente quando se progride na carreira até os níveis mais elevados.

A esse tipo de comportamento dá-se o nome de "efeito tesoura", numa referência à forma do gráfico em que duas curvas complementares (no caso do sexo, correspondentes a homens e mulheres) se afastam ou até se cruzam, lembrando uma tesoura aberta. O eventual cruzamento, necessariamente no nível de 50%, refletiria a inversão da predominância de um grupo sobre o outro. Mas mesmo quando não há cruzamento das curvas, o seu afastamento indica que a situação de disparidade se acentua com o tempo.

O efeito tesoura com relação aos sexos também ocorre na área da física, tanto no Brasil como no resto do mundo [1,2]. Mas, em que idade começa esse efeito? Qual é sua intensidade nas diferentes etapas da vida das mulheres?

Para medir a evolução do desempenho das meninas que se interessam por física durante o ensino fundamental e médio, utilizamos como indicador o número de premiações nas Olimpíadas Brasileiras de Física (OBF) ao longo dos anos de 2006 a 2015. As OBF, patrocinadas e organizadas com o apoio da Sociedade Brasileira de Física (SBF), ocorrem anualmente e alunos de escolas públicas e particulares entre o 8°ano do ensino fundamental (EF) e a 3° série do ensino médio (EM) podem participar. As premiações consistem em medalhas de ouro, prata e bronze, havendo também menções honrosas. Salientamos que os participantes que prestam os exames das OBF não necessariamente seguem carreiras científicas. Entretanto, essa participação indica um interesse por enfrentar e resolver desafios da física.

Para caracterizar a proporção de mulheres também nos diferentes estágios da formação e da carreira de física, analisamos os perfis dos bolsistas do CNPq (Conselho Nacional de Desenvolvimento Cientifico e Tecnológico) abrangendo a distribuição de bolsas desde os níveis de graduação (iniciação cientifica), pós-graduação (mestrado e doutorado) e de estímulo à produtividade científica (pesquisadores).

A Figura 1(a) apresenta o percentual de premiações na OBF (incluindo medalhas de ouro, prata e bronze, e também as menções honrosas) para alunas e alunos de cada ano escolar, sendo mostradas as médias, e o desvio padrão, computados sobre os anos desde 2006 a 2015. Como pode ser visto na Figura 1(b), para cada estágio escolar, os valores para os diferentes anos da década estudada são muito similares, sem grande dispersão, o que se reflete nas pequenas barras de erro representadas na Figura 1(a), e indica que a situação não tem mudado ao longo dessa década. O resultado evidencia um declínio no percentual de mulheres premiadas desde o 8° ano do EF até o 3° no EM, indicando que o efeito tesoura começa antes mesmo da decisão por uma carreira científica. Cabe mencionar que o mesmo declínio ocorre quando acompanhamos no tempo uma mesma turma. Além disso, na Figura 1(c) são exibidos os números totais, que mostram que não apenas o percentual de premiadas se reduz gradativamente, mas os números absolutos também decaem, enquanto os de meninos premiados aumentam. Portanto, o efeito tesoura não decorre simplesmente do caráter complementar dos números percentuais.

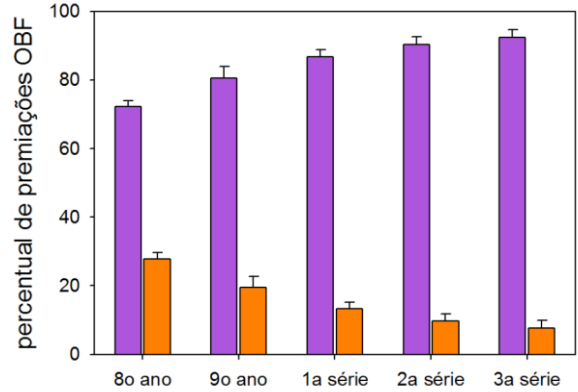
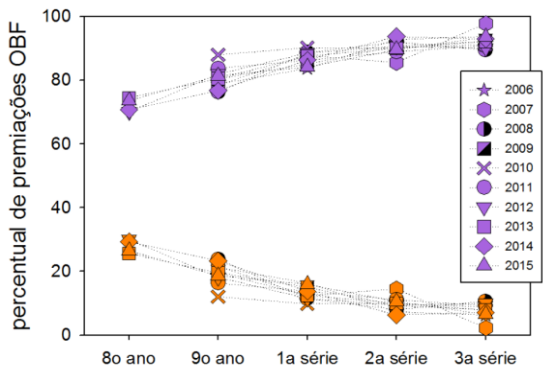
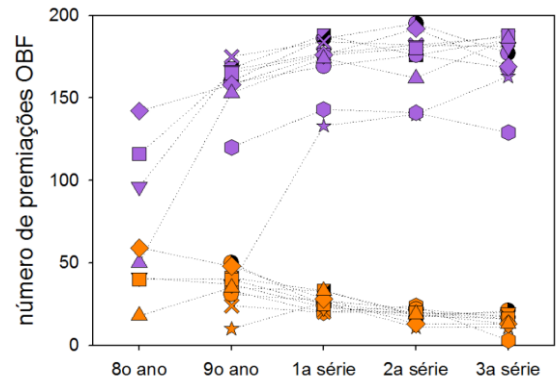

**Figura 1**: **Premiações na OBF:** (a) As colunas representam o percentual de mulheres (laranja) e homens (lilás) com premiações em cada ano escolar. Os valores apresentados correspondem à média calculada sobre os anos 2006-2015 (exceto de 2012-2015 para o 8º ano) e as barras de erro mostram o desvio padrão. (b) Percentuais para cada ano de 2006 a 2015 separadamente. (c) Números totais de premiações para os mesmos anos. Os dados foram extraídos da página web da SBF [4].

A Figura 2(a) ilustra que o mesmo efeito tesoura ocorre a partir do momento em que a pessoa decide seguir a carreira de física, através do número de bolsas do CNPq concedidas em diferentes estágios da carreira. Incluímos também as bolsas de iniciação científica júnior (ICj), destinadas a despertar vocação científica e incentivar talentos potenciais entre estudantes do ensino fundamental, médio e profissional da rede pública. Observa-se que nesta fase incipiente os percentuais de meninas e meninos são bastante próximos, com 45% de mulheres bolsistas. Ao ingressar na faculdade, os alunos podem pleitear bolsas de iniciação científica (IC), e se observa uma redução no percentual de meninas com relação à etapa anterior, que cai para 33%. Esta redução continua na pós-graduação - mestrado (M) e doutorado (D) – onde esse percentual cai a aproximadamente 21%. Segundo dados acessíveis na página do CNPq [3], os percentuais de bolsas para mulheres na física são sempre mais baixos do que os percentuais médios de bolsas distribuídas para mulheres de todas as áreas, que são de 59% para bolsas de IC, 52% para bolsas de M e 51% para D, sempre maiores do que os percentuais de homens.

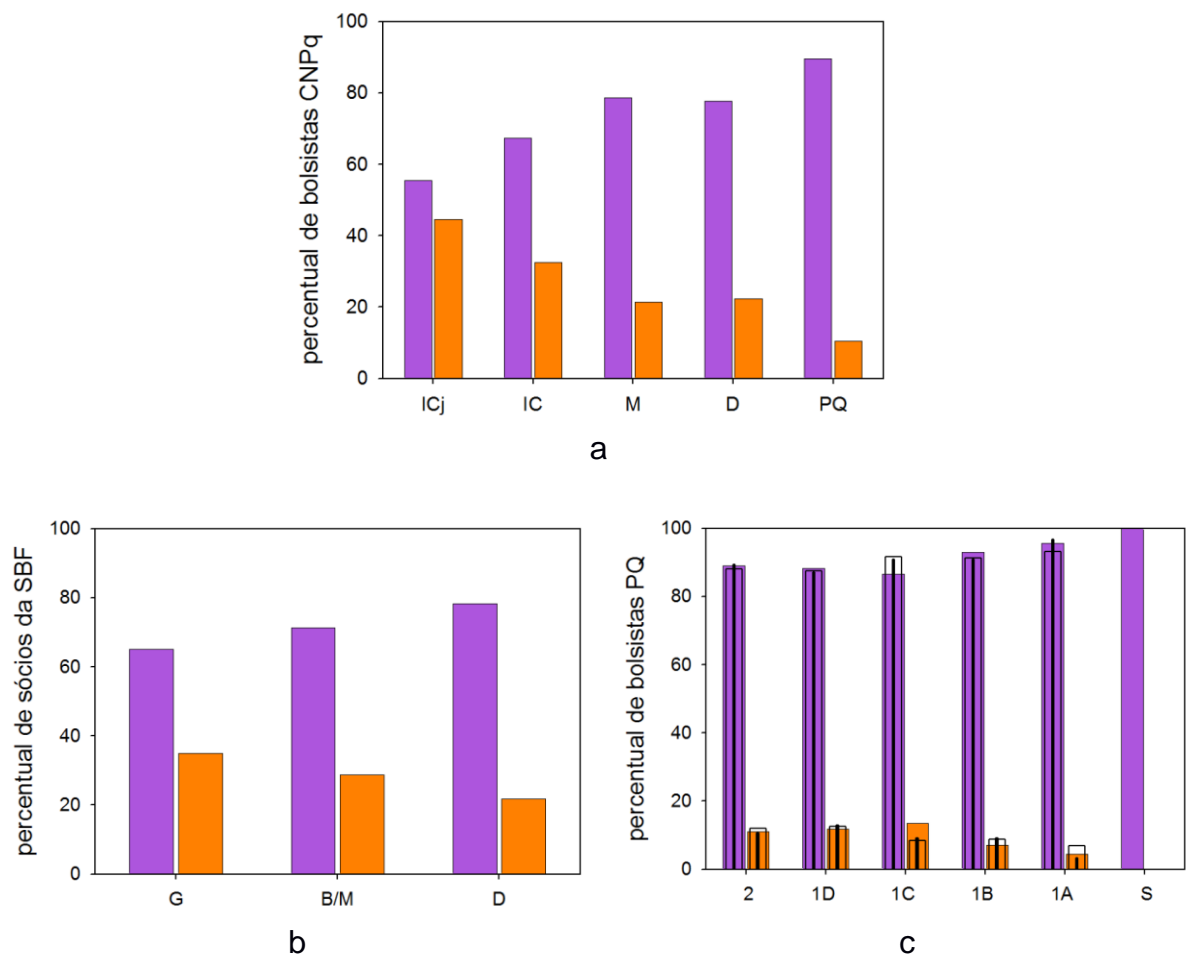

**Figura 2**: **Na universidade e na carreira:** (a) Percentual de bolsistas do CNPq mulheres (laranja) e homens (lilás): iniciação cientifica junior (ICj), iniciação cientifica (IC), mestrado (M), doutorado (D) e produtividade em pesquisa (PQ). (b) Percentual de mulheres (laranja) e homens (lilás) filiados à SBF: cursando a graduação (G), formados ou cursando o mestrado (B/M), doutores (D). Dados fornecidos pela SBF em 2017. (c) Percentual de bolsistas de produtividade de cada categoria. Dados extraídos da página web do CNPq em novembro de 2016. São mostrados também os dados extraídos de [5], correspondentes aos anos de 2005 (colunas pretas finas) e 2010 (colunas intermediárias).

Finalmente, ao se tornarem pesquisadores em uma instituição de ensino superior, doutores em física podem concorrer a uma bolsa de produtividade em pesquisa (PQ) oferecida pelo CNPq, bolsa esta destinada a valorizar a produção científica. A Figura 2(a) mostra que o percentual de doutoras que conquistam as bolsas PQ na área da física fica em torno de 10% (!), enquanto na média para todas as áreas atinge 36% [3]. Por um lado, estes dados mostram que, na média sobre as diversas áreas, existe um corte grande com relação aos percentuais em estágios anteriores da carreira e, por outro lado, que esse corte na física é mais acentuado que na média. Salientamos que esse retrato registrado em 2016 não melhorou em comparação com contagens semelhantes efetuadas em 2005 e 2010 [4], ao menos para a área da física.

O quadro que se reflete na distribuição de bolsas do CNPq nos diversos estágios também se manifesta na proporção de associados à SBF em diferentes categorias, como mostrado na Figura 2(b).

Mas será que o efeito tesoura cessa neste ponto? Podemos olhar em mais detalhe as bolsas PQ. Existem 6 níveis distintos, sendo o nível 2 o inicial, subindo para 1D, 1C, 1B, 1A, e Sênior, este último para pesquisadores que se destacaram por vários anos seguidos na sua área de atuação. Com base nos números mostrados acima, não é de se esperar que as mulheres sejam maioria em nenhum dos casos discutidos a seguir, mas o que vemos na Figura 2(c) são números muito inferiores em todas as categorias, e que, a partir do nível 1C, caem até chegar abaixo de 6% no nível mais alto da carreira científica (nível 1A), não havendo ainda mulheres na categoria Sênior, o que reflete uma situação histórica mais aguda do que a atual. Novamente, este retrato é essencialmente semelhante ao observado desde 2005 [5].

A ocorrência do efeito tesoura em etapas distintas da carreira nos leva a uma óbvia reflexão: quais são as causas deste fenômeno universal e robusto ao longo dos anos? Quais os fatores que impactam diferentemente as escolhas de homens e mulheres levando a números inicialmente inferiores de mulheres e à diminuição da participação das mulheres ao longo dos anos de estudos e carreira de física?

Uma das possíveis causas são os estereótipos e preconceitos de gênero aos quais as crianças são expostas desde muito jovens [6]. Alguns exemplos se encontram em filmes da Disney, nos quais a maioria das personagens femininas são princesas solitárias que não têm profissão e aguardam seus príncipes para darem sentido às suas vidas. Outros exemplos vem dos jogos de computadores para meninas (Jogos Online, [7]) onde os objetivos, em geral, são associados à limpeza de locais ou embelezamento pessoal, enquanto meninos são estimulados com jogos de montagem, robótica ou esportes de ação. Os impactos destes estereótipos têm consequências que já foram quantitativamente medidas, como mostra, por exemplo, um estudo publicado recentemente [8]. Este artigo aponta que a autoconfiança das meninas sofre uma mudança já aos 6 anos de idade. Para chegar a este resultado, os autores contam histórias de pessoas "muito inteligentes" – sem mencionar o gênero da pessoa – a crianças e perguntam a elas qual o gênero dos personagens recém apresentados. Aos 5 anos de idade, meninos e meninas associam tais personagens inteligentes ao seu próprio gênero. No entanto, apenas 2 anos mais tarde, quando o mesmo estudo é repetido com crianças de 7 anos de idade, ocorre uma importante mudança: a maioria das meninas associa um personagem inteligente ao gênero masculino. Os autores discutem as implicações dessa associaçao com relação à escolha de profissões: como matemática e ciências ditas "duras" são em geral consideradas difíceis, as meninas já quando muito jovens não se sentem aptas a seguir estas carreiras por não se considerarem capazes.

Outros estudos apontam que fatores como a falta de modelos de mulheres de sucesso e a existência de crenças ou estereótipos negativos sobre as suas habilidades e sobre as suas possibilidades de ascensão podem criar um sentimento de ''não pertencer'' e desestimular, assim, a participação em áreas específicas [9,10]. Cabe mencionar que o "preconceito inconsciente" de gênero se manifesta também na hora de avaliar candidatos, inclusive por parte das próprias mulheres [11] e que a participação destas nos processos seletivos é também desigual.

O primeiro passo para mudar essa realidade é adquirir consciência do fenômeno, reconhecer que é um problema, já que diversidade é essencial para a ciência e para qualquer atividade criativa e de inovação. O segundo passo é identificar os fatores que levam à disparidade de gênero, para finalmente poder interferir adequadamente nessa realidade. Uma tarefa neste sentido seria estimular o interesse das meninas por ciências exatas ainda antes da adolescência, apresentando para elas um cenário acolhedor, no qual as mulheres sintam que podem vir a fazer parte e serem bem sucedidas.

Estudos recentes [12] indicam não haver nas fases mais avançadas da carreira no Brasil, nenhum viés de gênero concreto no que concerne aos indicadores utilizados na concessão de bolsas PQ. As conclusões destes estudos devem servir de estímulo para que mais mulheres aptas a obter a bolsa por produtividade resolvam encaminhar suas solicitações ao CNPq. Cabe salientar porém que essa equidade foi observada apenas na última análise baseada nos dados de 2016. Em estudos anteiores, realizados em 2005 e 2010, ainda eram claramente perceptíveis diferenças que demonstravam uma exigência maior para mulheres do que para homens nos mesmos níveis de bolsas PQ.

Também esperamos que este texto seja útil para que cientistas das áreas ligadas ao estudo de gênero possam formular teses que ajudem a entender melhor as escolhas diferenciadas que, desde os ensinos fundamental e médio, afastam as meninas da física e de outras carreiras ligadas às ditas "ciências duras".

**Agradecimentos –** Agradecemos à Profa. Belita Koiller, presidente da SBF, pela sugestão de fazermos a análise dos dados da OBF, a Karina Buss (UFSC), Daniela Hiromi Okido, Sofia Guse (UFRGS), e Marlon Ramos (Unicamp) pela contagem de dados e aos demais membros do GTG-SBF, Andrea Simone Stucchi de Camargo, Antonio Gomes Filho e João Plascak por importantes sugestões.
## Referências

[1] E.B. Saitovitch, B.S. Lima, M.C. Barbosa, Mulheres na Física: uma análise quantitativa em Mulheres na Física (2015).

[2] Science policies in the European Union, Promoting excellence through mainstreaming gender equality, http://cordis.europa.eu/pub/improving/docs/g_wo_etan_en_199901.pdf, visitada em 16/08/2017.

[3] http://www.cnpq.br/web/guest/series-historicas, dados de 2014, visitada em 06/11/2016.

[4] http://www1.fisica.org.br/gt-genero/index.php/alguns-dados, visitada em 16/08/2017.

[5] P. Duarte, M. C.B. Barbosa e J.J. Arenzon, Produtividade em Pesquisa – CNPq, Física 2005-2010: uma análise comparativa, Instituto de Física – UFRGS.

[6] C.Brito, D.Pavani, P.Lima Jr., Meninas na Ciência: Atraindo jovens mulheres para carreiras de ciência e tecnologia, - Revista Gênero, v 16, p 33-50, (2005).

[7] Jogos online, http://jogosdemeninas.uol.com.br/, visitada em 18/08/2017.

[8] L. Bian, S-J. Leslie, A. Cimpian, Science 355, 389–391 (2017).

[9] S. Cheryan, S.A. Ziegler, A.K. Montoya, L. Jiang, (2016, October 10). Why Are Some STEM Fields More Gender Balanced Than Others? Psychological Bulletin. Advance online publication, doi: 10.1037/bul0000052.

[10] Why Europe's girls aren't studyng STEM, Microsoft Coorporation, http://www.voced.edu.au/content/ngv%3A76105, visitada em 11/07/2017.

[11] C. A. Moss-Racusina, J. F. Dovidio, V.L. Brescoll, M.J. Graham and J. Handelsman, Science faculty's subtle gender biases favor male students, PNAS 109, 41, 16474–16479, doi: 10.1073/pnas.1211286109.

[12] Menezes, D.P., Brito, C., Buss K., Anteneodo C., Bolsistas de produtividade em pesquisa em Física e Astronomia: análise quantitativa da produtividade científica de homens e mulheres, http://www1.fisica.org.br/gtgenero/images/arquivos/Apresentacoes_e_Textos/dados_CNPq_2016_vf.pdf, visitada em 11/07/2017.